# An Optimal Tunable Josephson Element for Quantum Computing

F. Chiarello, M. G. Castellano and G. Torrioli
*Istituto di Fotonica e Nanotecnologie – CNR, via del Cineto Romano 42, 00156 Rome, Italy*

S. Poletto
*Physikalisches Institut III, Universität Erlangen-Nürnberg, Erwin-Rommel-Str. 1, 91058 Erlangen, Germany*

C. Cosmelli
*Dip. Fisica, Università di Roma "La Sapienza", P.le A. Moro 2, 00185 Rome, Italy*

P. Carelli
*Ingegneria Elettrica, Università dell'Aquila, Monteluco di Roio, I-67040 L'Aquila, Italy*

D. V. Balashov, M. I. Khabipov and A. B. Zorin
*Physikalisch-Technische Bundesanstalt, Bundesallee 100, 38116 Braunschweig, Germany*

*We introduce a three-junction SQUID that can be effectively used as an optimal tunable element in Josephson quantum computing applications. This device can replace the simple dc SQUID generally used as tunable element in this kind of applications, with a series of advantages for the coherence time and for the tolerance to small errors. We study the device both theoretically and experimentally at 4.2 K, obtaining a good agreement between the results.*

Solid state superconducting devices are suitable candidates for the implementation of quantum computing, and different topologies of superconducting qubits, the basic elements of a quantum computer, have been realized and successfully tested[1-4]. All these systems are based on Josephson junctions, which are characterized by strongly nonlinear quantum behaviour. In many examples of existing qubits[5-9] a Josephson junction is replaced by a two-junction small interferometer (dc SQUID) used as tunable Josephson element (TJE). In the case of sufficiently small inductance the symmetric interferometer behaves approximately like a single junction with its main parameter - the critical current - modulated from a maximum value to a minimum (close to zero) by an externally applied control magnetic flux (fig.1a). This allows controlling and modifying the parameters of the qubit. In addition, there are also devices based on the TJE that are not qubits but are used for complementary tasks such as the qubit readout[10] and the controllable coupling of flux qubits[11,12].

Tunability is important for the fine adjustment of qubit parameters, or for having a flexible way to operate the qubit. On the other hand, each control acting on the qubit is a potential source of noise, for example through the control bias lines from the room temperature environment. For this reason the TJE is a further source of decoherence and this can be a severe problem for superconducting qubits.

In the so-called "Quantronium"[5] a successful strategy has been adopted to achieve a good compromise between ability to tune the device and its decoherence. The modulation curve of a dc SQUID critical current is flat in correspondence of the maximum, at zero flux bias. In this particular point (called optimal or "quiet" point) the system is insensitive (in the first order) to small variations of the flux bias, for example due to the incoming noise. The Quantronium is maintained at this point throughout the manipulation time (except during the readout) so that decoherence is strongly reduced. The same strategy can be used in different cases, but unfortunately the dc SQUID presents just one optimal point (apart for a periodic structure). If one chooses the optimal point strategy one can no longer tune the device and vice versa. In a general case, it may be required to switch between at least two distinct operational points, but only one of them (and not the other) can be an optimal point. For this reason it would be very important the use of a TJE that has two distinct optimal points and can be switched between them.

In this paper we show a tunable Josephson element (called optimal TJE, or O-TJE) with the described characteristic: the device behaves approximately like a Josephson junction, whose critical current can be modulated between a maximum and a minimum that are both optimal points (fig. 1b). The first advantage, as mentioned above, consists in a reduced sensitivity of the device to fluctuations around the optimal points. A second advantage consists in a very good tolerance to errors on the controls flux: a small error in this parameter setting is less disturbing in the case of fig. 1b with respect to the classical case of fig. 1a because of sufficient flatness of the characteristic at both optimal points.

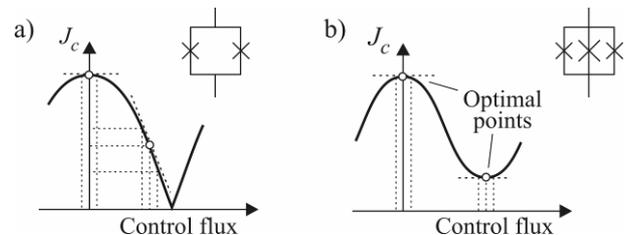

Fig. 1. (a) Modulation of the critical current in a standard low inductance dc SQUID (the Tunable Josephson Element). (b) Modulation of the critical current in a Optimal TJE, with indicated the two distinct optimal points.

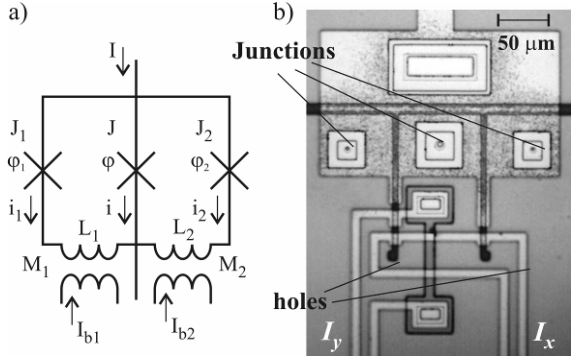

Fig. 2. (a) Scheme of the O-TJE considered in this paper. (b) Microphotograph of the tested device.

In the first part of this work we introduce the O-TJE, and study its theoretical behaviour. In the last part we consider the realization and the experimental characterization of a test device. The results are in agreement with the expected behaviour; in particular it can be verified the optimal behaviour by identifying the two distinct optimal points.

The O-TJE consists of three unshunted junctions in parallel (see references[13] for the description of similar devices used in different contests), with critical currents $J_1$, $J$, $J_2$ and capacitances $C_1$, $C$, $C_2$ respectively, inserted in two adjacent loops of inductances $L_1$ and $L_2$ (fig. 2a). Two distinct bias magnetic fluxes can be applied by means of two coils with currents $I_{b1}$ and $I_{b2}$, coupled to the loops with mutual inductances $M_1$ and $M_2$ respectively. In order to characterize the device behaviour we consider a current-biased stand-alone device (i.e. not inserted in a more complex system) and study the variation of its critical current versus the applied bias fluxes (fig. 1b), demonstrating its approximate equivalence to a single tunable Josephson junction.

The system dynamics is described by three degrees of freedom, the phase differences $\varphi_1$, $\varphi$ and $\varphi_2$ across the junctions $J_1$, $J$ and $J_2$, respectively. The equivalent potential can be derived from the scheme in fig. 2a, considering the sum of the three Josephson junctions energy contributions, plus the energy contributions of the two inductances, minus the work related to the ideal current source that generates the bias current $I$. By considering the relations between phases and currents in the loops, $\Phi_b(\varphi_1 - \varphi) + L_1 i_1 = M_1 I_{b1}$, $\Phi_b(\varphi - \varphi_2) - L_2 i_2 = M_2 I_{b2}$ and $I = i_1 + i + i_2$ (where $i_1$, $i$ and $i_2$ are the currents in the three branches, and $\Phi_b = \Phi_0/(2\pi)$ is the reduced flux quantum), one obtains:

$$U = -J_1 \Phi_b \cos\varphi_1 - J \Phi_b \cos\varphi - J_2 \Phi_b \cos\varphi_2 - I \Phi_b \varphi + \frac{\Phi_b^2}{2L_1}(\varphi - \varphi_1 + \theta_1)^2 + \frac{\Phi_b^2}{2L_2}(\varphi - \varphi_2 - \theta_2)^2 \quad (1)$$

where we use the reduced flux biases $\theta_1 = M_1 I_{b1}/\Phi_b$ and $\theta_2 = M_2 I_{b2}/\Phi_b$.

In the limit of negligible inductances ($L_1, L_2 \ll \Phi_b/J$, essential for the correct operation of the O-TJE) the last two terms in eq.1 (the inductive energy contributions) describe quasi-rigid bounding conditions that freeze two of the three degrees of freedom, so that in the zero order it is $\varphi_1 \cong \varphi + \theta_1$ and $\varphi_2 \cong \varphi - \theta_2$, and the potential becomes:

$$U \cong -J_1 \Phi_b \cos(\varphi + \theta_1) - J_2 \Phi_b \cos(\varphi - \theta_2) - J \Phi_b \cos\varphi - I \Phi_b \varphi, \quad (2)$$

This expression can be rearranged in a form that is equivalent to the potential of an effective single junction, plus a constant shift of the coordinate:

$$U \cong -J_c \Phi_b \cos(\varphi + \delta) - I \Phi_b \varphi, \quad (3)$$

where the modulation of the critical current $J_c$ and the phase shift $\delta$ are defined respectively as

$$J_c = \left[(J + J_1 \cos\theta_1 + J_2 \cos\theta_2)^2 + (J_1 \sin\theta_1 - J_2 \sin\theta_2)^2\right]^{1/2} \quad (4)$$

and

$$\delta = \arctan\left(\frac{J_1 \sin\theta_1 - J_2 \sin\theta_2}{J + J_1 \cos\theta_1 + J_2 \cos\theta_2}\right). \quad (5)$$

Note that in the case of a stand-alone TJE the constant phase shift $\delta$ has no important effects, but in more complex systems (for example whenever the TJE is inserted in a superconducting loop), $\delta$ must be taken into account.

Let us consider the simple and important case, used in this work, of a symmetric device (identical side junctions, $J_1 = J_2$), with a flux-bias provided by the same current $I_x = I_{b1} = I_{b2}$ (fig. 2b) circulating in identical coils with $M_1 = M_2 = M$, so that the phase biases are $\theta_1 = \theta_2 = \theta = M I_x / \Phi_b$. In this case the phase shift is $\delta = 0$, and the device behaves like a single Josephson junction with controllable critical current:

$$J_c(\theta) = J + 2J_1 \cos\theta \quad (6)$$

which oscillates between the maximum $J + 2J_1$ (at $\theta = 0$) and the minimum $J - 2J_1$ (at $\theta = \pi$), with a period of $2\pi$. The values $\theta = 0$ and $\pi$ correspond to the desired optimal, quiet points, where the first derivative of the critical current with respect to the bias is zero (fig. 1b). In order to tune the device, for example to compensate fabrication tolerances, we have introduced a second bias coil with current $I_y$ (shown in fig. 2b), mainly coupled to just the first loop (with mutual inductances $M_{y1} \gg M_{y2}$). The quiet-point strategy cannot be applied for this control but, since it is just a fixed dc-bias current, the noise contribution can be strongly reduced (for example it can be heavily filtered or, better, it can be provided by a stable superconducting trapping circuit[8]). The combined effect of the two controls gives the new reduced flux biases $\theta_{1,2} = M_{1,2} I_x / \Phi_b + M_{y1,2} I_y / \Phi_b$.

The appropriate choice of the device parameters, in particular of the $J_1/J$ ratio, together with the fine-tuning eventually operated by $I_y$ (if required), allows to tailor the device to the required application, from the qubit control[5-9] to the flux-coupling switching[11,12].

We have designed and fabricated applying the trilayer Nb/AlO$_x$/Nb technology [14] a series of chips, containing different O-TJEs with nominal parameters $J=30$ $\mu A$, $J_1=J_2=12$ $\mu A$, $L_1=L_2=7$ $pH$ (Fig. 2b). The characterization is performed at $T$ = 4.2 K, in a µ-metal shield, with R-C-R filters on the lines. The device is biased by a current ramp that causes a switching to the voltage state for some critical current value $J_c^*$ that is recorded by a computer acquisition system. This is repeated many times (100 – 10000) in order to obtain the statistical distribution of the switching current,, from which the effective critical current $J_c$ can be extrapolated, with standard techniques[15]. The procedure is repeated for different values of the flux bias currents $I_x$ and $I_y$ in order to obtain the experimental critical current vs. fluxes characteristic (solid constant-level lines in the contour plot of Fig.3). The result can be compared with the theoretical prediction given in eqs. 4 and 5 (dashed constant-level lines in the contour plot in Fig.3). The simple matching of peaks positions allows to determine $M_1 = 6.8\,pH \approx M_2 = 6.6\,pH$, and $M_{y1} = 12.4\,pH > M_{y2} = 1.7\,pH$. Once determined these values it is possible to fit the data, obtaining $J = 31.0\,\mu A$ and $J_1 = J_2 = 12.5\,\mu A$, in agreement with the expected values. A small discrepancy is visible for small values of the critical current; it is the analogue of the non-zero modulation of the critical current in a simple dc SQUID[10], and it is due to the effect of the small but non vanishing inductances $L_1$ and $L_2$. Fig.4 shows the modulation of the critical current with $I_x$ for $I_y$ fixed to -15.5 µA, and one can notice the distinct optimal points that can be used for the quiet operation of the O-TJE. The varying amplitude of the peaks is a spurious effect due to the small asymmetry between $M_1$ and $M_2$, visible also in the not perfect horizontal alignment of the peaks in Fig.3. Except for the discrepancy at low $J_c$ discussed above, the agreement with theory is good, so that the device can be effectively used as a flux-controlled tunable Josephson junction with two distinct optimal points.

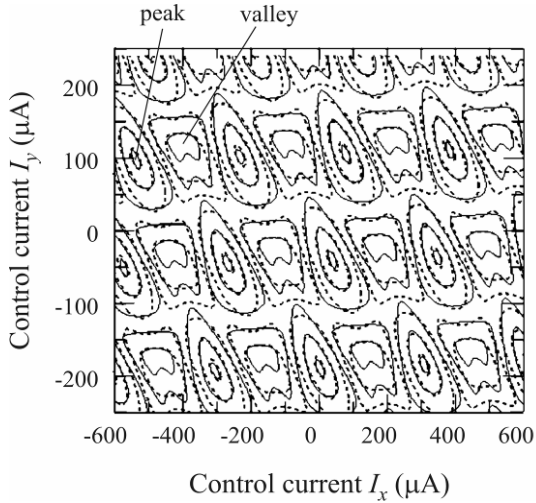

Fig. 3. Contour plots of the critical current vs the flux current controls $I_x$ and $I_y$ (solid constant-level lines for the experimental data, dashed level lines for the fit). The contours correspond to critical current values $J_c$ = 55 µA, 50 µA, 40 µA, 30 µA, 20 µA, respectively (from peaks to valleys).

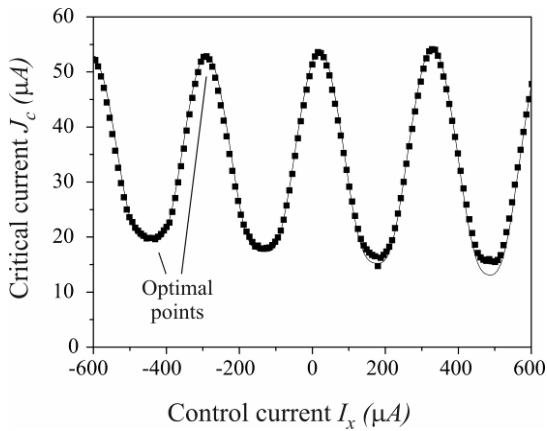

Fig. 4. Modulation of the critical current $J_c$ vs the current $I_x$ for fixed $I_y = -15.5\,\mu A$.

To conclude, we introduced a three-junctions SQUID (the O-TJE) that can be used as tunable element in Josephson quantum computing applications instead of the simple dc SQUID, with possible improvement of the coherence time. The device behaviour has been investigated both theoretically and experimentally, obtaining a good agreement. The future work will consider both the insertion of the device in more complex systems and the study of the effects of the non-zero inductances on the dynamics.

This work was supported by the European Commission, contract FP6-502807 (RSFQubit), and by the CNR RSTL program.